\documentclass[12pt,a4paper]{article}
\usepackage{a4wide}
\usepackage{latexsym}
\usepackage{epsf}
\usepackage{amssymb}
\begin{document}
\def\be{\begin{equation}}
\def\ee{\end{equation}}
\def\bea{\begin{eqnarray}}
\def\eea{\end{eqnarray}}

\def\pd{\partial}
\def\a{\alpha}
\def\b{\beta}
\def\g{\gamma}
\def\d{\delta}
\def\m{\mu}
\def\n{\nu}
\def\t{\tau}
\def\l{\lambda}

\def\s{\sigma}
\def\e{\epsilon}
\def\scri{\mathcal{J}}
\def\cM{\mathcal{M}}
\def\tcM{\tilde{\mathcal{M}}}
\def\RR{\mathbb{R}}
%%%%%%%%%%%%%%%%%%%%%%%%%%%%%%%%%%%%%%%%%%%%%%%%%%%%%%%%%%%%%%%%%%%%%%

\hyphenation{re-pa-ra-me-tri-za-tion}
\hyphenation{trans-for-ma-tions}

%%%%%%%%%%%%%%%%%%%%%%%%%%%%%%%%%%%%%%%%%%%%%%%%%%%%%%%%%%%%%%%%%%%%%%

\begin{flushright}
IFT-UAM/CSIC-99-03\\
hep-th/9902012\\
\end{flushright}

\vspace{1cm}

\begin{center}

{\bf\Large Non-Critical Confining Strings and the Renormalization Group}

\vspace{.5cm}

{\bf Enrique \'Alvarez}${}^{\diamondsuit,\clubsuit}$
\footnote{E-mail: {\tt enrique.alvarez@uam.es}}
{\bf and C\'esar G\'omez}${}^{\diamondsuit,\spadesuit}$
\footnote{E-mail: {\tt iffgomez@roca.csic.es}} \\
\vspace{.3cm}

\vskip 0.4cm

${}^{\diamondsuit}$\ {\it Instituto de F\'{\i}sica Te\'orica, C-XVI,
  Universidad Aut\'onoma de Madrid \\
  E-28049-Madrid, Spain}\footnote{Unidad de Investigaci\'on Asociada
  al Centro de F\'{\i}sica Miguel Catal\'an (C.S.I.C.)}

\vskip 0.2cm

${}^{\clubsuit}$\ {\it Departamento de F\'{\i}sica Te\'orica, C-XI,
  Universidad Aut\'onoma de Madrid \\
  E-28049-Madrid, Spain}

\vskip 0.2cm

${}^{\spadesuit}$\ {\it I.M.A.F.F., C.S.I.C., Calle de Serrano 113\\ 
E-28006-Madrid, Spain}

\vskip 1cm

%%%%%%%%%%%%%%%%%%%%%%%%%%%%%%%%%%%%%%%%%%%%%%%%%%%%%%%%%%%%%%%%%%%%%%

{\bf Abstract}

String vacua for non critical strings satisfying the requirements of
 Zig-Zag invariance are constructed. The Liouville mode is shown to 
play the r\^ole of scale in the Renormalization Group operation.
Differences and similarities with the D-brane near horizon
approach to non supersymmetric gauge theories are discussed as well.

\end{center}
%%%%%%%%%%%%%%%%%%%%%%%%%%%%%%%%%%%%%%%%%%%%%%%%%%%%%%%%%%%%%%%%%%%%%%

\begin{quote}

\end{quote}

%%%%%%%%%%%%%%%%%%%%%%%%%%%%%%%%%%%%%%%%%%%%%%%%%%%%%%%%%%%%%%%%%%%%%%

\newpage
%%%%%%%%%%%%%%%%%%%%%%%%%%%%%%%%%%%%%%%%%%%%%%%%%%%%%%%%%%%%%%%%%%%%%%

\setcounter{page}{1}
\setcounter{footnote}{1}

\section{Introduction}
%%%%%%%%%%%%%%%%%%%%%%%%%%%%%%%%%%%%%%%%%%%%%%%%%%%%%%%%%%%%%%

At present we can consider two different philosophies to approach the
problem of defining a string representation of non abelian four
dimensional gauge theories. \footnote{Both can be considered as different
attempts to implement {\em holographic} ideas \cite{'th}. }
One approach is based on Maldacena's
conjecture \cite{malda} and on the so called 
AdS-CFT correspondence \cite{gkp}\cite{witten}.  In this
approach we start with a stack of D-3 branes in type IIB string theory
and we perform a near horizon limit introducing a new blow-up variable
that can be identified with the scale of the gauge theory. In its
strong version Maldacena's conjecture stablishes the equivalence
between N=4 supersymmetric Yang Mills theory in flat Minkowski
space-time and Type IIB strings in $AdS_5\times S_5$
 with a non vanishing Ramond-Ramond (R-R)
five form background and Dirichlet boundary conditions to be defined
at Penrose's conformal infinity \cite{penrose}.
 The main difficulty with this approach
is of course its extension to gauge theories with non vanishing beta
function. This extension requires working with stacks of
D-branes reproducing on their world volume non conformal theories, and
performing on them a similar bulk decoupling near horizon limit. In
this direction, two different
possibilities have recently been suggested. 
One due to Witten \cite{witten1} 
starts by considering the near
horizon limit of a stack of $M_5$ branes in M-Theory. The resulting six
dimensional conformal field theory is subsequently compactified to
four dimensions, breaking explicitly supersymmetry. 
The other
possibility due to Klebanov and Tseytlin \cite{kt} consists
 in working from the
begining with non supersymmetric Dixon-Harvey \cite{dh} 
type 0(B) strings. 
The
difference between both approaches mainly refers to the relation
between the string scale and the breaking of supersymmetry scale.

\par
In type 0(B) strings one starts with a superstring with world sheet
supersymmetry that contains closed string tachyons. The idea is to use
the dynamics of these tachyons (that contrary to what happens in the
bosonic case now  have non vanishing amplitudes for even number of
external legs only) to induce an effective central charge that will
automatically produce a non trivial dependence of the dilaton field on
the radial coordinate. \footnote{Given the fact that the RR repulsion is 
twice as strong as the 
NSNS attraction,
in \cite{kt} one is forced to hope that the r\^ole 
of tachyons is to
first allow the stacking of branes by forming a bound state; so 
tightly bound, 
in fact, that it would remain so even when the tachyon condenses.}
\par

It would be this dependence the one to be
identified with the renormalization group equation for the non
supersymmentric gauge theory defined on the stack of D3 branes.
\par
 It is interesting to observe that in both cases we define a non 
conformal theory by breaking completely supresymmetry.

\par
The other approach, due to Polyakov \cite{polyakov}\cite{ago}
\cite{ks} \cite{dkt}\cite{fm}
 to the string representation of
non abelian gauge theories is based on the so called Zig-Zag (ZZ)
invariance of pure Yang Mills Wilson loops \cite{libro}. The main idea is to
impose on open string amplitudes invariance under generic
reparametrizations that are not necessarily  orientation preserving
diffeomorphisms. This requirement implies a priori both the absence of
open string tachyons, which can be achieved by working with a world
sheet supersymmetric string theory \cite{polyakov}
\cite{migdal}\cite{ago} and an appropiate GSO projection,
and the truncation of the open string spectrum to pure massless vector
bosons. It is due to this truncation that ZZ invariant strings
are naturally describing pure gauge theories without performing any
extra bulk decoupling limit. In order to describe a four dimensional
gauge theory in the ZZ approach we should start with a non
critical open string theory in four dimensional flat Minkowski
space-time. Since this theory is anomalous with respect to world-sheet
Weyl rescalings it is necessary to add a Liouville extra mode, that is
effectively acting as an extra dimension. A ZZ invariant
background will correspond to a five dimensional space-time metric of
the type:
\be
ds^2 = a(\phi)^2 dx_{\parallel}^2 + d\phi^2
\ee
(where $x_{\parallel}\in M_4$, the Minkowski space in which the putative
gauge theory lives) and Dirichlet boundary conditions on the Liouville 
mode on the ZZ horizon:
\be
a(\phi^{\star}) = 0.
\ee

If we find such a solution we can  try to reproduce from it
the renormalization group equation for the pure gauge theory defined
by the open string sector. 
\footnote{ The other possibility, consisting in working with
$a=\infty$ can be interpreted as some T-dual of (1.2) \cite{polyakov}}.
Since we are now working in a non critical
string in flat Minkowski space-time, the dilaton field will depend on
the Liouville coordinate and we can use this dependence to reproduce
the running of the coupling constant. In the non critical case the
identification of the extra Liouville coordinate with the scale of the
theory defined by the non critical string in four dimensions is quite
natural. In fact the very meaning of the Liouville mode is to
compensate two dimensional Weyl rescalings which in flat space are
equivalent to dilatations of the four dimensional space-time
coordinate: 
\be 
x_{\parallel}\rightarrow \lambda x_{\parallel}. 
\ee 
In
this sense the renormalization group equation for the physical open string
amplitudes of the zig zag invariant theory would be formally defined
as follows:
\be
0 = \frac{d}{d\lambda}A = (\delta x_{\parallel}
\frac{\delta}{\delta x_{\parallel}} + 
\delta \phi \frac{\delta}{\delta \phi}) A.
\ee

It is important to stress the differences and similarities between
the ZZ approach and the type 0(B) string approach. In both cases
the reason for the running of the dilaton field appears as a
consequence of the existence of an effective central charge;
their origin however is completly different. In the Type 0(B) 
case the 
effective central charge stems from the existence of a tachyon
condensate while in the non critical case it is a consequence of
working off criticality.
\par
Let us now explain in what precise way
ZZ invariance forces us to work off criticality. If we work
Maldacena's near horizon limit, in the critical case, Penrose's
conformal infinity will possess the general structure of $M_4\times S_5$
 and thus
Dirichlet open string boundary conditions will be imposed on $M_4\times S_5$
However ZZ invariant boundary conditions require a truncation of all
Kaluza Klein modes in order to enforce Dirichlet boundary conditions strictly
on $M_4$ , working in that way in a non critical string. The main 
difference ,however, has its origin in the geometry of the
bulk decoupling. In the case of a stack of D 3 branes we need,
following Maldacena, to work out the near horizon limit while in the
ZZ non critical case the bulk decoupling is explicitely done by
demanding Dirichlet boundary conditions on the ZZ horizon, i.e
the region of space time where the four dimensional metric in the string
frame vanishes. Moreover in the non critical case as pointed out above
the Liouville coordinate appears as a natural scale of the theory.
\footnote{A different way of interpreting the blow-up coordinate stems
from a T-duality $R\rightarrow \frac{\a'}{R}$, by taking the double
limit $\a'\rightarrow 0$ and $R\rightarrow 0$, with $\frac{\a'}{R}\equiv u$.
Notice that the resulting theory is a full fledged string.
In this sense, this double limit defines a transformation between 
different string vacua.}
\par

In order to understand properly the dynamical r\^ole played by the
Liouville field it would be important to introduce the concept of
{\em Liouville frame} \cite{gkp}. Let us do it using for that 
Maldacena's example, where it is natural  to identify the blow up coordinate
with a Liouville coordinate. This identification can be understood as
follows. Let us parametrize the transversal space to a D 3 brane in
type IIB in terms of the Euler angles in $S_5$ and the radius of such
sphere i.e using polar coordinates. Let us now forget formally about
the radial coordinate and pretend that we are working in a non
critical case where the transversal space is simply five dimensional
and with the topology of the sphere. We know that this string theory
interpreted as non critical generates an extra Liouville coordinate.
By definition Liouville frame will mean to consider the string theory
in the new coordinates such that one of them is Liouville itself. 
Very likely
the topology will change by going to this new frame where what we get
naturally is a transversal space of the type considered in the bulk
near horizon limit, namely $S_5\times \mathbb{R}$, where $\mathbb{R}$
 refers now to the Liouville
coordinate. 
\footnote{The combined result of the two operations implies first, to forget
about some coordinates, and moreover, replacing them by a Liouville mode
which will in general define a non-trivial transformation between 
string vacua}
\par
In summary it looks that the Liouville frame is the
natural one to capture the bulk decoupling physics. In defining the
Liouville frame it is important the type of coordinates we choose at
the begining, in Maldacena's example polar coordinates. The ones that
we keep are the ones related with global $\mathcal{R}$ symmetries.  In Liouville
frame the Liouville coordinate is always playing the r\^ole of a scale
of the non critical theory, this fact being one of the reasons why 
this frame is
the appropiated one to describe bulk decoupling limits.

\par
In the same way as it happens for type 0(B) strings the ZZ
approach is directly working in the no space-time supersymmetric
regime. This is in principle due to the fact that ZZ invariant
reparametrizations for open string amplitudes already truncate the
open string spectrum to the pure NS sector (cf. \cite{ago}). 
The vertex operators for
fermionic R states are manifestly not ZZ invariant, since they
involve, in any picture, the path integration of the space-time spinor
field (which lives in a frame related to a non-zero vierbein).
Due to this fact and the already mentioned absence of open
string tachyons in  ZZ invariant strings we are in the ZZ
approach forced to work with GSO projections that are not space time
supersymmetric. A more general question we can address at this point
is whether space time supersymmetry is or not consistent with non
criticality, where by that we generically mean non critical central
charge. If we reduce the problem to the case of flat space-time a
temptative answer can be given. In fact if we consider four
dimensional non critical strings and we add a super Liouville mode, it
seems difficult to get a space-time GSO proyection in five dimensions
since there are not superstrings in five dimensions. 
\footnote{That is, it is not known how to implement $\kappa$-symmetry there.}
A different
situation will appears in the case we start with a three dimensional
non critical string. In that case we can get after adding super
Liouville a GSO in four dimensions consistent with space time
supersymmetry.

\par
Concerning space-time supersymmetry let us just make the following
general comment. If by some dynamical procedure some condensation takes
place in a critical string theory that generates an effective central
extension, we conjecture that such a dynamical process will induce
dynamical breaking of space-time supersymmetry. For instance if we
think in a type IIB string any condensation leading to an effective
central charge will inmediately break the $SL(2,\mathbb{Z})$
 duality invariance
and consequently space-time supersymmetry.
\par

In this paper we will work in the ZZ approach for non critical
strings. Next we shall review our main results. First we consider the
simplest case with no tachyon background and non RR background and we
look for ZZ invariant solutions. We find a unique solution that
degenerates to the trivial one if the missing central charge is zero.
This solution induces a running for the dilaton field that is of the
asymptotically free type, with the coupling decreasing in the
ultraviolet. Moreover we find a phase transition with a conformal
point. Since in order to achieve ZZ invariance we need to
decouple R open string states it is natural to turn on the tachyon. We
do that qualitatively and we find that the tachyon expectation value
can move the location of the conformal point leaving us with a pure
asymptotically free theory. In all our analysis we work in the string
frame and we don't add any form of R R background.
\par

From our analysis we can suggest a generalization of Maldacena's
conjecture. In the ZZ framework we can say that five dimensional
superstring vacua , enjoying a four dimensional ZZ horizon and
satisfying Dirichlet boundary conditions on it are completly
equivalent to pure four dimensional gauge theories. This generalization
of Maldacena's conjecture replaces critical string by non critical
ones, near horizon limit by ZZ invariance, Penrose's conformal
infinity by ZZ horizon and the standard AdS-CFT correspondence by
imposing Dirichlet boundary conditions at the ZZ horizon. The
main difficulty, as already mentioned, with the ZZ invariant
approach is that due to the fact that R fermionic vertex operators are
not ZZ invariant we should deal with the presence of a closed
string tachyon, and that, since in this approach we are not dealing
with any stack of branes, the information about the specific gauge
group should be included somehow in the particular way to fix the
tachyon vacuum expectation value. 
\par
One possibility related with the
particular examples we work out in this paper consists in starting with
open superstrings in four dimensions with the standard GSO projection
i.e. without any form of tachyon and to consider the solution to the
tachyonless beta function equations. Most likely this solution does not
admit space time supersymmetry and if we enforce Dirichlet boundary
conditions on the ZZ horizon then the spectrum of the open
string sector is reduced to the one corresponding to a pure gauge theory. 

Surprisingly enough we
do not get the renormalization group behavior of a pure Yang Mills
theory indicating that something is missing in the whole argument.

%%%%%%%%%%%%%%%%%%%%%%%%%%%%%%%%%%%%%%%%%%%%%%%%%%%%%%%%%%%%%%%%%%%%%
\section{Non-Critical Confining Strings and the Renormalization Group}
%%%%%%%%%%%%%%%%%%%%%%%%%%%%%%%%%%%%%%%%%%%%%%%%%%%%%%%%%%%%%%%%%%%%%%%%%

In the search for confining strings we are lead to 
 consider a {\em non-critical string}, where the would-be holographic
coordinate is now interpreted as a Liouville field. The corresponding
two-dimensional action (including the necessary Dilaton and Tachyon 
fields) is:
\be
S = \int d^2 z [ (\pd\phi)^2 + a^2 (\phi) (\pd x_{\parallel})^2 + 
R^{(2)}\Phi(\phi) + T(\phi)]
\ee

\par
In order to study the renormalization group equations of the Yang-Mills
theory living on the horizon, we need to examine how ordinary dilatations of the
Minkowskian coordinates are implemented in the confining string framework.
\bea
&&ds^2_{\parallel}\rightarrow \lambda^{-1}ds^2_{\parallel}\nonumber \\
&&a(\phi)\rightarrow \lambda a(\phi)
\eea
(this corresponds to translations of the Liouville field in the simplest case 
). It is to be stressed that {\em the ZZ Horizon itself remains invariant
under these transformations}. 
\par
In the flat space case, this means
\bea
&&\delta x_{\parallel} = \epsilon x_{\parallel}\nonumber \\
&&\delta(log{a}) = -\epsilon
\eea
This suggests that there is an identification between the logarithmic
dilatation on the horizon (which is identified up to a sign 
with a logarithmic variation
of the renormalization scale) and the logarithmic variation of the varying 
string tension itself:
\be
\frac{\d x_{\parallel}}{x_{\parallel}}\equiv - \frac{\d \mu}{\mu}
= - \frac{\d a(\phi)}{a(\phi)}
\ee
This then implies unambiguosly that the Infrared (IR) region is on the 
horizon itself ($a = 0$); whereas the Ultraviolet (UV) region is located
on the boundary ($ a\rightarrow\infty$).Curiously enough, although our framework
is not holographic {\it sensu stricto}, the geometrical identifications of the
energy scales coincides with the holographic situation \cite{sw}.

\par
The transformation of the Liouville field  does not then in general
leave invariant the kinetic term, nor the dilaton.
The variation of the action is:
\bea
\delta S &=& \int d^2 z [-\pd_a\phi \pd^a(\frac{a}{a'})\epsilon  + (T'(\phi)
+ \Phi'(\phi) R^{(2)})
\frac{a}{a'}\epsilon ] \nonumber\\
 &\sim &\int d^2 z \frac{a}{a'}\epsilon ( \pd_a\pd^a \phi + (T'(\phi) 
+ R^{(2)} \Phi^{\prime}
(\phi)) )
\eea

\par
From this point of view, the universal behavior of the dilaton under
Weyl transformations would be:
\be
\frac{\delta\Phi}{\delta\epsilon}= - 
\frac{\delta \Phi}{\delta\phi}\frac{a(\phi )}{a'(\phi)}
\ee

\par
The condition for persistence of the horizon in the Einstein metric is
\be
lim_{a\rightarrow 0}\quad e^{\frac{2}{3}\Phi} a^2 = 0
\ee
This means that the divergence of the coupling constant when reaching the horizon
must be smaller than $a^{-3}$.
\footnote{Under this circumstances, we can equally well (by redefining 
the Liouville
coordinate) interpret the above {\em ansatz} as representing the metric 
in the Einstein frame, which is sometimes convenient when studying the 
equations of motion. Those derive from the action principle
\par
\bea
S_E &=& \frac{1}{16\pi G_5}\int d^5 x \sqrt{g_E} [ R_E + \frac{1}{3} 
(\nabla \Phi)^2\nonumber\\
&+&\frac{10}{3\alpha'}e^{-\frac{2}{3}\Phi} + \frac{1}{6}((\nabla T)^2 - m^2
e^{-\frac{2}{3}\Phi}T^2)]\nonumber
\eea
 The tachyon mass is given by $m^2 = -\frac{2}{\a '}$.
\par
For example, when the tachyon vanishes,
the Einstein equations of motion then reduce to:
\bea
&&\Phi '' + 4 \Phi '\frac{a'}{a} = - \frac{20}{3\a'} e ^{-\frac{2}{3}\Phi}
\nonumber\\
&&4\frac{a''}{a} +\frac{1}{3}(\Phi')^2-\frac{2}{a^2}(2 a'' + 3 (a')^2)
+\frac{1}{6}(\Phi')^2 + \frac{5}{3\a'} e^{-\frac{2}{3}\Phi}=0\nonumber\\
&&a a'' + 3 (a')^2 - 4 a'' - 6 (a')^2 +\frac{a^2}{6}(\Phi')^2 +\frac{5 a^2}{\a'}
e^{-\frac{2}{3}\Phi}=0\nonumber
\eea
}
\par
Owing to our identification above (8), between the renormalization
group scale $\mu$ ans the {\em string frame} metric coefficient $a(\phi)$, 
we shall continue working in the string frame for ease of the physical
 interpretation.
\par
The Weyl anomaly coefficients \cite{cp} in the said string frame read
\bea
\beta^{*}_{\m\n}&=&\a'R_{\m\n} -\a'\nabla_{\m}\nabla_{\n}\Phi\nonumber\\
\beta^{*}_{\Phi}&=& = \a'\nabla^2 \Phi + \frac{2}{3}(D-10) + \a'(\nabla \Phi)^2\nonumber\\
\beta^{*}_{T}&=&\a'\nabla^2 T  - 4 T + \a'\nabla_{\m}\Phi\nabla^{\m} T
\eea
We have taken into account the central charge defect 
$-\frac{2}{3\a '}(D-10)$, where we use the value $10$ because the model ,as 
argued above,  enjoys two-dimensional
supersymmetry.
\par
In the presence of a tachyon field, there are extra terms (formally of
order $\a'^2$), so that the complete equations read
\footnote{Those equations coincide with the ones in a footnote in page 12
of \cite{kt} with the identification $\Phi_{kt} \equiv -\frac{1}{2} \Phi$}
\bea
R_{\m\n} -\nabla_{\m}\nabla_{\n}\Phi
+\frac{1}{4}\nabla_{\m}T\nabla_{\n}T &=& 0\nonumber\\
- \nabla^2 \Phi + c_0  - (\nabla \Phi)^2
- \frac{m^2}{4} T^2 &=& 0\nonumber\\
\nabla^2 T  - m^2 T + \nabla_{\m}\Phi\nabla^{\m} T &=& 0
\eea
and where now  the tachyon mass is given by $m^2\equiv -\frac{D-2}{4\a'}$ and
the central charge defect is $c_0\equiv -\frac{D-10}{\a'}$.
%%%%%%%%%%%%%%%%%%%%%%%%%%%%%%%%%%%%%%%%%%%%%%%%%%%%%%%%%%%%%%%%%%%%%
\subsection{Tachyonless Backgrounds}
%%%%%%%%%%%%%%%%%%%%%%%%%%%%%%%%%%%%%%%%%%%%%%%%%%%%%%%%%%%%%%%%%%%%%%

Assuming that the tachyon (through the interaction with the RR backgrounds
or otherwise) develops a vacuum expectation value, in a first approximation
all its effect will be to change the numerical value of the central charge
defect $c_0$ . Putting by simplicity the 
tachyon field to zero and assuming the {\em radial ansatz} $\pd_{\mu}\Phi = 0$
where $x^{\mu}\in M_{\parallel}$, yields:
\bea
&&R_{\phi\phi} = - 4 \frac{a''}{a} = \Phi''\nonumber\\
&&R_{\m\n}= -(a a'' + 3 (a')^2)\delta_{\m\n} = - \Gamma_{\m\n}^{\phi}\Phi' = 
a a' \Phi'\delta_{\m\n}\nonumber\\
&&\Phi'' + 4 \Phi'\frac{a'}{a} - c_0 + (\Phi')^2 = 0
\eea
where $f'\equiv \frac{d f}{d\phi}$.
\par
There are now two subcases to consider. When $a' = 0$, then necessarily
\be
\Phi'' = 0
\ee
and 
\be
\Phi' = c_0^{1/2}
\ee
This is the well-known {\em linear dilaton} solution.
\par
When $a'\neq 0$, on the other hand,a linear combination of the above equations 
can be expressed solely in terms of the function $a(\phi)$:
\be
- 2 \frac{a''}{a} + (\frac{a''}{a'})^2 - 3 (\frac{a'}{a})^2 - c_0 = 0
\ee
which can be integrated by the substitution
\be
a \equiv e^{\int u}
\ee
This yields
\be
- 4 u^2 + (\frac{u'}{u})^2 - c_0 = 0
\ee
which can be easily integrated, giving:
\be
u = \frac{\sqrt{c_0 (t_0^2 - c_0)}}{2[\g_0 \cosh{\g_0(\phi - \phi_0)} - t_0
\sinh{\g_0(\phi -\phi_0)}]}
\ee
where $\g_0^2 \equiv c_0$, and when $\phi = \phi_0$, $u = u_0\equiv 
\frac{1}{2}\sqrt{
t_0^2 - c_0}$.
\par
Given the fact that necessarily $\g_0 < t_0$, there is a singularity
at the value of the Liouville coordinate given by
\be
\phi_{sing}= \phi_0 + \frac{1}{\g_0} th^{- 1}(\g_0 /t_0)
\ee
(Only in the singular limiting case when $ t_0 = \g_0$ is $\phi_{sing}$
pushed towards $\phi_{sing} = \infty $).
\par
Integrating again (21) leads to:
\be
a(\phi)= \a_0\sqrt{\frac{1+\lambda e^{\g_o(\phi - \phi_0)}}{1 - 
\lambda e^{\g_o(\phi - \phi_0)}}}
\ee
where
\be
\lambda^2 = \frac{t_0 - \g_0}{t_0 + \g_0}
\ee
and when $\phi = \phi_0$, $a = a_0 \equiv \a_0 \sqrt{\frac{1 + \lambda}{1 - 
\lambda}}$. Again, for positive values of the parameter $\lambda$, 
the function explodes when $\phi = \phi_{sing}$.
\par

%%%%%%%%%%%%%%%%%%%%%%%%%%%%%%%%%%%%%%%%%%%%%%%%%%%%%%%%
\subsubsection{A ZZ Invariant solution}
%%%%%%%%%%%%%%%%%%%%%%%%%%%%%%%%%%%%%%%%%%%%%%%%%%%%%%%%%

When $t_0=\infty$,  taking the valuation $\lambda = - 1$, and changing $\phi$ 
by $-\phi$ (which is a symmetry of the equations (14)),
\footnote{This last step is neccessary in order to get a solution which
{\em starts}, rather that {\em ends} at the horizon} we get:
\be
a(\phi)= \a_0\sqrt{\frac{1- e^{- \g_o(\phi - \phi_0)}}{1 + 
e^{- \g_o(\phi - \phi_0)}}}
\ee
which is the only ZZ invariant solution in the whole family, that
being in itself a remarkable fact.
\par
A curious thing is that solution flattens itself down (i.e., reduces to 
the trivial one $a\equiv 0$) when the central charge defect vanishes, that
is $c_0 = 0$.
\par
This solution starts at $a=0$ when $\phi= \phi_0\equiv \phi^{*}$, and grows
monotonically until it reaches the asymptotic value $a = \a_0$ (which
is arbitrary). This implies that, unless $\a_0=\infty$, it is not
self (T)-dual; that is, there is no region in the spacetime ({\em boundary})
with $a(\phi)=\infty$. This will have physicically important 
consequences, because, as we shall see, one is led to identify 
the Ultraviolet (UV) with the boundary, and the Infrared (IR) 
with the ZZ horizon. 
\par
Incidentally, the ZZ invariant solution starting of the horizon, and the
$\phi$-reversal, ending on it, would both be in a mutual T-dual relationship
were it not for the dilaton field \cite{aagl}.

%%%%%%%%%%%%%%%%%%%%%%%%%%%%%%%%%%%%%%%%%%%%%%%%%%%%%%%%%%%%%
\subsubsection{$g_{YM}(\mu )$ from the Dilaton}
%%%%%%%%%%%%%%%%%%%%%%%%%%%%%%%%%%%%%%%%%%%%%%%%%%%%%%%%%%%%

We would like to  interpret  
$ \pi^{3/2} g_s e^{\Phi} \equiv g_{YM}^2$ as the coupling constant of the 
putative 
gauge theory
living on the ZZ Horizon \cite{bachas}; but which coupling constant , 
bare or renormalized,
and if the latter, at which scale?
\par
Apparently, the only possible thing we can identify the dilaton with is with
the {\em running coupling constant}, and the sense of the running is
provided by our previous identification in (8) of the string implementation
of the Yang-Mills scale transformations.
\par
On general grounds, even before
integrating, we can write for our putative $\b$-function
\be
a\frac{d e^{\Phi}}{d a} = - (\frac{a a''}{(a')^2}+ 3 )e^{\Phi}\nonumber
\ee
Plugging there the former results from (4.25) one gets:
\be
e^{\Phi} = e^{\Phi_0}(\frac{a_0}{a})^6\frac{1 - (\a_0 / a_0)^4}
{1-(\a_0 / a)^4}
\ee
It can be easily checked that this is a decreasing function of $a$ as long as
$a > \a_0$. \footnote{This actually covers the whole domain of the Louville 
variable in the non zz-invariant situation when $\lambda >0$
because then $\a_0 < a_0$ always.}
\par
In the ZZ-invariant case, $e^{\Phi_0}$ diverges in such a way that
\be
e^{\Phi}\sim \frac{1}{a^6(\frac{\a_0^4}{a^4}- 1)}
\ee
This clearly decreases as a function of $a$ up to a given value of $a$,
namely $a_{\min}\equiv 3^{- 1/4}\a_0$, after which point it starts to
increase whithout bounds.\footnote{
Please note that owing to our identification (8) between the 
renormalization
group scale $\mu$ and the conformal factor $a$, we do not have any 
freedom in choosing the coordinate upon which the dilaton depends. 
This contrasts with
the situation in most other treatments.}
\par
The physical meaning of the turn-over point seems to stem from the fact
that it has horizontal tangent, that is, $\beta(g^{\star}) 
= 0$; it is a {\em conformally
invariant fixed point} .
\par
Incidentally, it is not difficult to show that in the vicinity of $g^{\star}$
the beta function reads
\be
\beta\sim - \frac{\a_0 3^{1/4}}{2} g^3 \, \sqrt{\frac{2}{3} - \frac{\sqrt{3}}
{\a_0^6 g^2}}
\ee
\par
On the other hand, even in the Asymptotically Free (AF) regime, the dependence
with the putative $\mu$ is {\em not} logarithmic.
The generic (not ZZ-invariant) solutions
 ($\lambda >0 $), although they 
are  asymptotically free (AF) in the whole allowed domain, also
lack logarithmic dependence.
\par

It is plain that once the function $a(\phi)$ is known, the dilaton 
can be easily extracted 
from the second equation of (14).
This leads at once in the generic case ($\lambda >0$) to
\footnote{We have explicitly verified that it also solves
the first equation of (14), which was only used as a substitution up to now}
\be
\Phi - \Phi_0 = \log \frac{[1-\lambda e^{\gamma_0(\phi - \phi_0)}]^3}
{e^{\gamma_0(\phi - \phi_0)} + \lambda e^{2 \gamma_0(\phi - \phi_0)}}
\frac{1+\lambda}{1-\lambda}
\ee
and for the ZZ-invariant solution to
\be
\Phi\sim 3\, \log( 1 + e ^{-\g_0 (\phi - \phi_0)})- 
\log(e ^{-\g_0 (\phi - \phi_0)} - e ^{- 2\g_0 (\phi - \phi_0)})
\ee
From here we can easily determine that the AF regime ends up at
\be
\phi_{min} - \phi_0 = \g_0^{-1} \log (2 + \sqrt{3})
\ee

%%%%%%%%%%%%%%%%%%%%%%%%%%%%%%%%%%%%%%%%%%%%%%%%%%%%%%%%%%%%%%%%%%%
\section{The Physics of Tachyon Condensates}
%%%%%%%%%%%%%%%%%%%%%%%%%%%%%%%%%%%%%%%%%%%%%%%%%%%%%%%%%%%%%%%%%%

It is a well known fact \cite{cp} that in the presence of 
arbitrary NS massless
condensates the spacetime effective action vanishes on shell because it is
precisely proportional to the dilaton beta function, $\beta_{\Phi}$. 
\par
When there is a tachyon condensate, it is easy to check that on shell
\be
S_{eff}\equiv - 2 \int \sqrt{G} d^D x\, (2 c_0 -\frac{m^2}{2} T^2)e^{-2\Phi}
\ee
In the range of spacetime dimensions between $D=2$ and $D=10$, then, the fact
that the tachyon mass is {\em negative} precisely enforces
\be
S_{eff}\leq 0
\ee
thus implementing Zamolodchikov's c-theorem in the present context, which
physically means \cite{libro} that
\be
 C = D - 10 + S_{eff}
\ee

%%%%%%%%%%%%%%%%%%%%%%%%%%%%%%%%%%%%%%%%%%%%%%%%%%%%%%%%%%%%%%%%%%
\subsection{Tachyonful Backgrounds}
%%%%%%%%%%%%%%%%%%%%%%%%%%%%%%%%%%%%%%%%%%%%%%%%%%%%%%%%%%%%%%%%%%

The full equations (13) with the Tachyon turned on \footnote
{There are other terms, neglected here, which are of the same formal order
in the $\a'$-expansion.}read:
\bea
&&R_{\phi\phi} = - 4 \frac{a''}{a} = \Phi'' + \frac{1}{4} (T')^2\nonumber\\
&&R_{\m\n}= -(a a'' + 3 (a')^2)\delta_{\m\n} = - \Gamma_{\m\n}^{\phi}\Phi' = 
a a' \Phi'\delta_{\m\n}\nonumber\\
&&\Phi'' + 4 \Phi'\frac{a'}{a} - c_0 + (\Phi')^2 + \frac{m^2}{4} T^2 = 0
\nonumber\\
&&T'' + 4 T'\frac{a'}{a} - m^2 T + \Phi' T' = 0
\eea

The complete set of equations is quite difficult to solve exactly. What we 
can do instead is to examine the behavior of a tachyon in the tachyonless
background of the preceding section, and then study how this tachyon back-reacts
on the said background.
\par

First of all, we can trace from (35) the origin of the turnover point
in the behavior of the dilaton as a function of $a$
to the vanishing of the Heaviside function $\theta (a a'' + 3 (a')^2)$.
\par
Now, the analogous to our previous equation (17) in the presence of a tachyon
is:
\be
- 2 \frac{a''}{a} + (\frac{a''}{a'})^2 - 3 (\frac{a'}{a})^2 - c_0 
- \frac{(T')^2 - m^2 T^2}{4} = 0
\ee
This means that on the tachyonless geometry the quantity
\be
\frac{a a''}{(a')^2} + 3 = 4 \pm \sqrt{(e^{\gamma_0 (\phi -\phi_0)} + 
e^{-\gamma_0 (\phi -\phi_0)})^2 + (e^{\gamma_0 (\phi -\phi_0)} - 
e^{-\gamma_0 (\phi -\phi_0)})^2 \frac {(T')^2 - m^2 T^2}{4\gamma_0^2}}
\ee
(Where it should always be taken the negative valuation of the square root).
\par
When the tachyon vanishes this gives:
\be
\frac{a a''}{(a')^2} + 3 = 4 - e^{\gamma_0 (\phi -\phi_0)} - 
e^{-\gamma_0 (\phi -\phi_0)}
\ee
which passes through a zero precisely at the value $\phi_{min}$ quoted above
after equation (27).
\par
It is clear now that the effect of a {\em free} tachyon is always to approach
the turnover point to the origin, because the extra term in the square root
is positive definite (this is a purely {\em tachyonic} effect).
\par
The only possibility for the tachyon dynamics to push the turnover point 
towards $\phi = \infty$ would be that a positive definite potential
$V(T)$
is generated, in such a way that the term $m^2 T^2$ is replaced by
$m^2 T^2 + V(T)$.
\par
To be specific, if we define the quantity
\be
\rho\equiv \frac{V(T) + m^2 T^2 - (T')^2}{4\gamma_0^2}
\ee
then the condition that the Heaviside function $\theta = 1$ implies that 
$(1-\rho)(z^2 + z^{-2})+ 2 (1+\rho) <16$, $\forall z\in (0,\infty)$.
This is clearly only possible when $\rho = 1$, which is the same as:
\be
V(T) = 4 c_0 - m^2 T^2 + (T')^2
\ee
which, in the tachyonless background reduces to
\be
V(T) = 4 c_0
\ee

\par
Our present understanding of the dynamics of non-critical confining
strings does not allow us to gauge what are the odds for such a potential
to be generated in the present context.

%%%%%%%%%%%%%%%%%%%%%%%%%%%%%%%%%%%%%%%%%%%%%%%%%%%%%%%%%%%%%%%%%%%%
\subsection{Backreaction}
%%%%%%%%%%%%%%%%%%%%%%%%%%%%%%%%%%%%%%%%%%%%%%%%%%%%%%%%%%%%%%%%%%

If we plug this value for the tachyon potential back in our previous
equation (36) we get the simple equation
\be
a a'' = - (a')^2
\ee
whose general solution is of the form
\be
a = a_0 \sqrt{1 + 2 u_0 (\phi - \phi_0)}
\ee
Again, there is only one ZZ-invariant solution in the family, namely
\be
a\sim (\phi - \phi_0)^{1/2}
\ee
Using (27), this gives the behavior of the dilaton as:
\be
e^{\Phi}\sim (\phi - \phi_0)^{-5/4}.
\ee

\par
This is again AF in the whole allowed domain, although the dependence
with the variable we have argued before to be the correct
implementation of the renormalization group scale $\mu$ in the present
context, namely, $a$ itself, is not logarithmic but rather \be
g_{YM}^2\sim \mu^{-5/2}.  \ee 

It is interesting to notice that the
renormalization group we get is of a power law type, similar to the
one considered in the accelerated unification. \cite{ddgb} This behavior is
typical of a full fleged five dimensional theory, pointing out to a
potential problem in the ZZ scenario.
\par

%%%%%%%%%%%%%%%%%%%%%%%%%%%%%%%%%%%%%%%%%%%%%%%%%%%%%%%%%%%%%%%%%%%%%%%%%%
\section{Conclusions}
%%%%%%%%%%%%%%%%%%%%%%%%%%%%%%%%%%%%%%%%%%%%%%%%%%%%%%%%%%%%%%%%%%%%%%%%%%

In summary there are two different ways to perform bulk decoupling
limits or in other words to define string vacua equivalent to non
abelian gauge theories, namely near horizon limit of an stack of D-branes
and ZZ invariant solutions for non critical strings. ZZ
invariance is manifestly no space time supersymmetric and non critical
and allow us to read the evolution of the renormalization group
directly from the Liouville field dependence of the dilaton field.
Bulk decoupling for non conformal field theories with partial
supersymmetry breaking will require some different procedures.

%%%%%%%%%%%%%%%%%%%%%%%%%%%%%%%%%%%%%%%%%%%%%%%%%%%%%%%%%%%%%%%%%%%%
\section*{Acknowledgments}
We thank A. Delgado for pointing out reference \cite{ddgb}.
This work ~~has been partially supported by the
European Union TMR program FMRX-CT96-0012 {\sl Integrability,
  Non-perturbative Effects, and Symmetry in Quantum Field Theory} and
by the Spanish grant AEN96-1655.  The work of E.A.~has also been
supported by the European Union TMR program ERBFMRX-CT96-0090 {\sl 
Beyond the Standard model} 
 and  the Spanish grant  AEN96-1664.

%%%%%%%%%%%%%%%%%%%%%%%%%%%%%%%%%%%%%%%%%%%%%%%%%%%%%%%%%%%%%%%%%%%%%%

\appendix

%%%%%%%%%%%%%%%%%%%%%%%%%%%%%%%%%%%%%%%%%%%%%%%%%%%%%%%%%%%%%%%%%%%%%%

\end{document}